\documentclass[twocolumn,superscriptaddress]{revtex4}
\usepackage[colorlinks=true,linkcolor=blue, citecolor=blue]{hyperref}
\usepackage{graphicx}
\usepackage{subfigure}
\usepackage{dcolumn}

\begin{document}
	
	\title{Sensitivity of anisotropic flow in Pb+Pb collisions to the neutron skin of $^{208}$Pb at energies available at the CERN Large Hadron Collider}
	\author{Xin-Li Zhao}
	\affiliation{School of Physics, Faculty of Basic Sciences, University of Shanghai for Science and Technology, Shanghai 200093, China}
	\affiliation{Key Laboratory of Nuclear Physics and Ion-beam Application (MOE), Institute of Modern Physics, Fudan University, Shanghai 200433, China}
	\affiliation{Shanghai Research Center for Theoretical Nuclear Physics, NSFC and Fudan University, Shanghai 200438, China}
	\author{Xin-Yi Xie}
	\affiliation{School of Physics, Faculty of Basic Sciences, University of Shanghai for Science and Technology, Shanghai 200093, China}
	\author{Yuan Li}
	\affiliation{Key Laboratory of Nuclear Physics and Ion-beam Application (MOE), Institute of Modern Physics, Fudan University, Shanghai 200433, China}
	\affiliation{Shanghai Research Center for Theoretical Nuclear Physics, NSFC and Fudan University, Shanghai 200438, China}
	\author{Guo-Liang Ma}
	\email{glma@fudan.edu.cn}
	\affiliation{Key Laboratory of Nuclear Physics and Ion-beam Application (MOE), Institute of Modern Physics, Fudan University, Shanghai 200433, China}
	\affiliation{Shanghai Research Center for Theoretical Nuclear Physics, NSFC and Fudan University, Shanghai 200438, China}

	\begin{abstract}
		
		We study the sensitivity of anisotropic flow in Pb+Pb collisions at $\sqrt{s_{\mathrm{NN}}}=5.02$~TeV to the neutron skin of $^{208}$Pb using the improved string-melting version of a multiphase transport (AMPT) model. By varying the neutron surface diffuseness, we trace the impact of several neutron-skin scenarios from the initial nuclear density profiles and eccentricities to final-state charged-particle production and flow cumulants. A systematic response is observed in both the initial eccentricities and the anisotropic flow, indicating that the neutron-skin-induced geometric effects can survive the full transport evolution. A model-to-data comparison with ALICE measurements disfavors the most compact and most diffuse density profiles within the tested AMPT setup, while the reference and moderate-skin configurations give comparable descriptions. Their similarity reflects a geometric degeneracy in large Pb+Pb collision systems, where the inclusive anisotropic flow is dominated by the effective collision geometry and nuclear size, making it only weakly sensitive to the detailed structure of the neutron surface. These results clarify both the sensitivity and the limitations of using anisotropic flow in relativistic heavy-ion collisions to probe neutron-skin effects.
		
	\end{abstract}

	\maketitle
	
	\section{\label{sec:intro} Introduction}
	
	Relativistic heavy-ion collisions are primarily used to study strongly interacting matter under extreme conditions. Over the past decades, measurements of bulk particle production and anisotropic flow have established a coherent picture of the collective expansion of the quark-gluon plasma (QGP)~\cite{Ollitrault:1992bk,Heinz:2013th,Chen:2024aom}. At the same time, the spatial distributions of nucleons in the colliding nuclei determine the fluctuating initial geometry of the produced medium and may therefore leave observable imprints on final-state collective observables. This connection has motivated extensive studies of nuclear deformation, triaxiality, clustering, and other nuclear-structure effects in relativistic collisions~\cite{Ma:2022dbh,Wang:2023rpd,Wibowo:2018gjg,Gorbunov:2022bzi,Jia:2022ozr,Zhang:2024vkh,STAR:2025elk,STAR:2024wgy,Giacalone:2021udy}. Collisions of isobaric systems, such as Ru+Ru and Zr+Zr, have demonstrated measurable sensitivity of flow observables to nuclear deformation and density distributions~\cite{Wang:2023yis,Zhao:2019crj,Zhao:2022grq,Zhang:2021kxj,Jia:2022qgl}, while collisions of oxygen and neon nuclei at the BNL Relativistic Heavy Ion Collider (RHIC)~\cite{Huang:2023viw} and the CERN Large Hadron
	Collider (LHC)~\cite{ALICE:2025luc,CMS:2025tga,ATLAS:2025nnt} have opened new opportunities for exploring cluster and shape effects in small and intermediate systems. Nevertheless, extracting quantitative nuclear-structure information from relativistic collisions is a model-dependent inverse problem, because the measured observables depend not only on the ground-state nuclear density but also on entropy deposition, centrality selection, partonic and hadronic transport, and the final collective response. Recent critical discussions have therefore emphasized the importance of distinguishing robust geometric sensitivity from model-dependent nuclear-structure extraction~\cite{Dobaczewski:2025rdi}. In this sense, heavy-ion collisions provide a complementary, but not model-independent, probe of nuclear structure.
	
	A particularly important property of neutron-rich heavy nuclei is the neutron-skin thickness of $^{208}$Pb, $\Delta r_{np}=\sqrt{\langle r_n^2\rangle}-\sqrt{\langle r_p^2\rangle}$, which characterizes the difference between the neutron and proton root-mean-square radii. The neutron skin is correlated with the density dependence of the nuclear symmetry energy and the equation of state of neutron-rich matter, although the quantitative connection depends on the underlying nuclear many-body framework~\cite{Roca-Maza:2011qcr,Tsang:2012se,Reed:2021nqk,Sammarruca:2022rcl,Burgio:2024vjc,Pihan:2025pep,Suzuki:2022mow}. A broad range of electromagnetic and hadronic probes has been used to constrain the neutron distribution of $^{208}$Pb, and existing determinations still exhibit significant spread~\cite{PREX:2021umo,Roca-Maza:2013mla,Malace:2010ft,Tagami:2023jkj,Mantysaari:2024uwn,Zhang:2021jwh}. The PREX-II parity-violating electron-scattering experiment reported a relatively large neutron-skin thickness for $^{208}$Pb, $\Delta r_{np}\approx0.283$~fm~\cite{PREX:2021umo}. In contrast, the CREX experiment measured the weak-charge distribution of $^{48}$Ca and found a comparatively small neutron skin for that nucleus~\cite{CREX:2022kgg}. Simultaneously describing the PREX-II and CREX results has proved challenging for many nuclear energy-density functionals and microscopic calculations, and has stimulated extensive discussions on the interpretation of parity-violating measurements, symmetry-energy constraints, and astrophysical observations~\cite{Reinhard:2021utv,Hu:2021trw,Sotani:2023qfe,Essick:2021kjb,Zhang:2022bni,Liu:2023pav,Ding:2024xxu,Yue:2021yfx,Huan:2024kfs,Giacalone:2023cet}. As emphasized in Ref.~\cite{Reinhard:2022inh}, the current experimental tension should be interpreted with caution before being converted into precise constraints on neutron skins or the symmetry energy. Independent and complementary probes of neutron-rich nuclear surfaces are thus crucial. Future parity-violating measurements, including the planned MREX experiment at MESA, will provide important new insights.
	
	Relativistic Pb+Pb collisions offer one possible complementary approach because the proton and neutron density profiles of $^{208}$Pb enter the initial matter distribution of the collision system. Modifications of the neutron-rich surface can change the participant density, the initial eccentricities, and the event-by-event geometry fluctuations that subsequently generate anisotropic flow. Previous studies over a wide range of collision energies have shown that isospin-differential observables, such as charged-pion ratios, neutron-to-proton ratios, or mirror-nucleus yield and flow ratios, can provide more direct sensitivity to isospin dynamics and neutron-rich surfaces~\cite{Wei:2013sfa,Helenius:2016dsk,Hartnack:2018sih,Li:2019kkh}. In contrast, most bulk observables measured at LHC midrapidity are charge inclusive and are expected to respond mainly to the total matter geometry rather than directly to isospin transport. At LHC energies, Ref.~\cite{Giacalone:2023cet} has already used Pb+Pb data in a Bayesian analysis within a different dynamical framework to extract the neutron skin of $^{208}$Pb with quantified uncertainties. Such studies demonstrate the potential of collider data, while also showing that the inferred neutron skin can depend on the assumed nuclear-density parametrization, prior information, and medium-response model. It is therefore useful to investigate the same physics from an independent microscopic transport perspective and to clarify which aspects of the neutron-density modification are actually transmitted to final-state flow.
	
	The goal of the present work is complementary to Bayesian parameter extraction. We perform a forward-model sensitivity study within the improved string-melting version of a multiphase transport (AMPT) model, which microscopically describes the evolution from initial particle production through partonic scattering, quark coalescence, and hadronic interactions. We focus on three questions. First, can the neutron-skin-induced geometric modifications of $^{208}$Pb survive the full multiphase transport evolution and remain visible in final-state anisotropic flow? Second, how do different flow observables, in particular two- and four-particle cumulants, respond to the tested neutron-skin scenarios? Third, to what extent do current charge-inclusive Pb+Pb observables discriminate among different neutron-density profiles, and where does a geometric degeneracy arise between configurations with similar effective matter geometry? 
	Accordingly, the present analysis is intended as a model-dependent study of sensitivity, dynamical transmission, and scenario discrimination rather than as a model-independent posterior determination of $\Delta r_{np}$.
	
	The paper is organized as follows. In Sec.~\ref{sec:model}, we introduce the improved AMPT model and the neutron-skin configurations of ${}^{208}$Pb. In Sec.~\ref{sec:results}, we present and discuss our results on initial-state eccentricities and anisotropic flow, and compare them with ALICE measurements in Pb+Pb collisions. Finally, conclusions are drawn in Sec.~\ref{sec:sum}.

	\section{\label{sec:model} Improved AMPT model and neutron-skin configurations}
	
	In this work, the string-melting version of the AMPT model (AMPT-SM)~\cite{Lin:2004en} is employed to simulate Pb+Pb collisions. The AMPT-SM framework provides a microscopic description of the full collision dynamics, including the generation of initial conditions by the Heavy Ion Jet Interaction Generator (HIJING), partonic scatterings described by Zhang's Parton Cascade (ZPC), hadronization via quark coalescence, and subsequent hadronic interactions modeled by A Relativistic Transport (ART). We use the improved version of AMPT-SM, which incorporates several significant improvements, such as an improved quark-coalescence mechanism~\cite{He:2017tla}, modern parton distribution functions for the free proton, impact-parameter-dependent nuclear shadowing, and an updated treatment of heavy-flavor production~\cite{Zhang:2019utb,Zhang:2021vvp,Zhang:2022fum}. Recent studies have shown that these improvements lead to a more consistent description of particle yields, transverse-momentum spectra, and anisotropic flow over a wide range of collision systems and energies~\cite{Zhang:2019utb,Zhang:2023xkw,Lin:2021mdn,Zheng:2019alz}. The updated framework therefore provides a reliable basis for investigating how initial geometric features are transmitted to the anisotropic flow.
	
	The spatial distributions of protons and neutrons in the ${}^{208}$Pb nucleus are modeled using the Woods-Saxon parametrization,
	\begin{equation}
		\rho(r)=\frac{\rho_0}{1+\exp\!\left[(r-R_0)/a\right]},
		\label{eq:ws}
	\end{equation}
	where $\rho_0$ is the saturation density, $R_0$ the nuclear radius, and $a$ the surface diffuseness parameter. Neutron-skin effects are incorporated by assigning different Woods--Saxon parameters to protons and neutrons. The parameter sets used in this work are summarized in Tabble.~\ref{tab:rnp}, corresponding to neutron-skin thicknesses $\Delta r_{np} = -0.19,\ 0,\ 0.16,\ 0.37,$ and $0.74~\mathrm{fm}$. The configuration with $\Delta r_{np}=0$ is taken as a reference density profile from the commonly used Woods-Saxon parametrization~\cite{Loizides:2017ack}. The other configurations are generated by varying the neutron diffuseness parameter $a_n$ while keeping $R_n$ and the proton density parameters fixed. We have also checked variations of the neutron radius parameter $R_n$ within the same transport setup, and found that the resulting changes in the flow observables considered here are much weaker. We therefore focus on the neutron surface-diffuseness variation, which provides a clearer test of how modifications of the neutron-rich surface are propagated to final-state anisotropic flow.
	
	\begin{table}[t]
		\caption{Woods-Saxon parameters for the proton and neutron density distributions of the $^{208}$Pb configurations for different neutron-skin thicknesses $\Delta r_{np}$, with $\Delta r_{np}=0$ taken as the reference configuration. All values are in fm.}
		\renewcommand\arraystretch{1.5}
		\label{tab:rnp}
		\begin{tabular}{cccc|c}
			\hline
			\hline
			~~~~$a_{n}~~~~$ & ~~~~$R_{n}$~~~~ & ~~~~$a_{p}$~~~~ & ~~~~$R_{p}$~~~~ & ~~~~$\Delta r_{np}$~~~~ \\ \hline
			0.20  & 6.69  & 0.447 & 6.68  & -0.19 \\
			0.549 & 6.624 & 0.549 & 6.624 & 0~\cite{Loizides:2017ack}     \\
			0.56  & 6.69  & 0.447 & 6.68  & 0.16~\cite{Nijs:2023yab}  \\
			0.70  & 6.69  & 0.447 & 6.68  & 0.37  \\
			0.90  & 6.69  & 0.447 & 6.68  & 0.74  \\ \hline\hline
		\end{tabular}
	\end{table}
	
\begin{figure}
		\begin{minipage}[c]{0.5\textwidth}
			\subfigure{\includegraphics[width=0.8\textwidth]{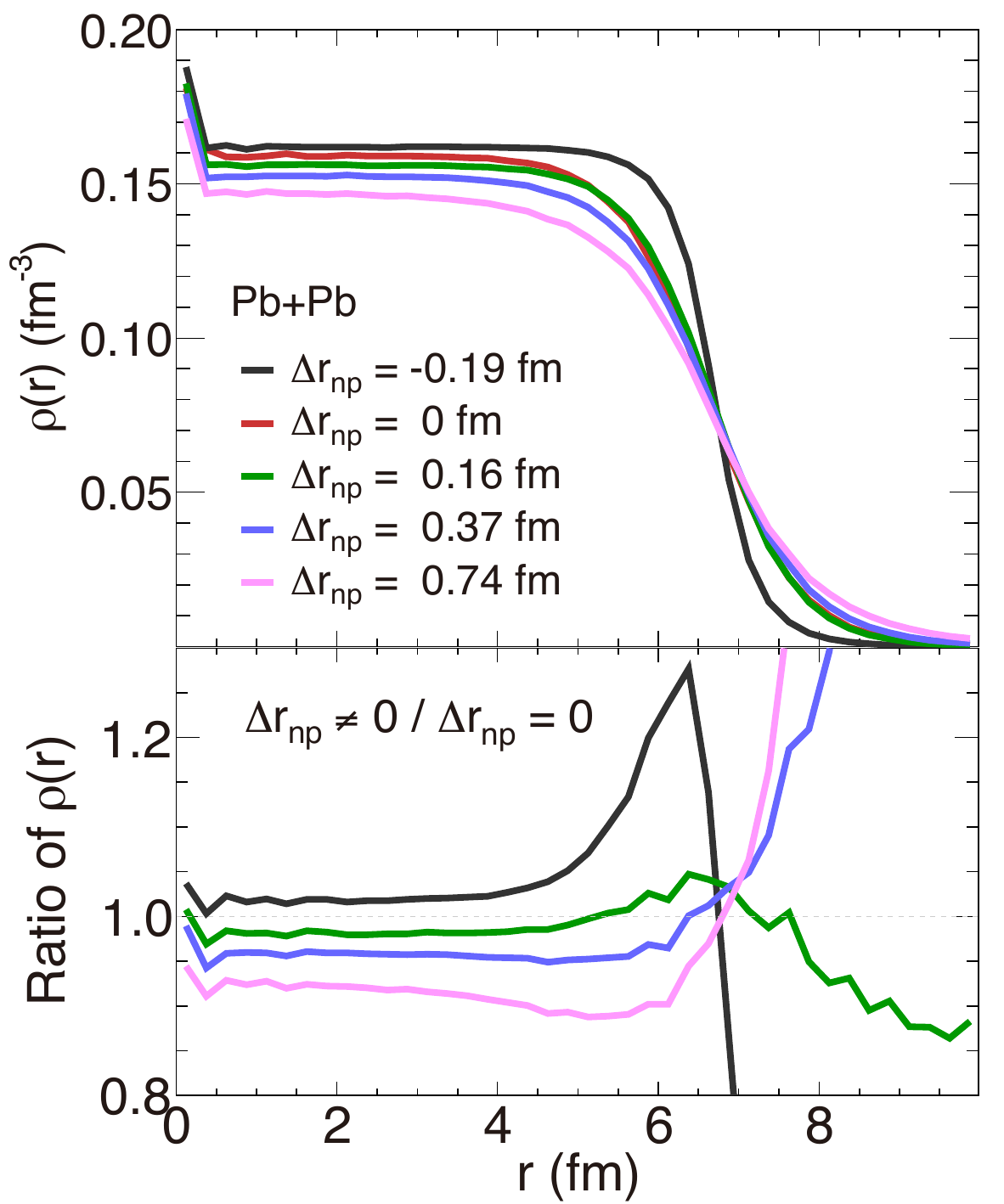}}
		\end{minipage}
		\caption{Upper panel: The nucleon density distributions of $^{208}$Pb in $\rm Pb+Pb$ collisions with different $\Delta r_{np}$. Lower panel: The ratio of the nucleon density distributions of $^{208}$Pb with $\Delta r_{np}$ relative to the reference configuration with $\Delta r_{np}=0$.}
		\label{fig:rho}
	\end{figure}

	Figure~\ref{fig:rho} illustrates the nucleon density $\rho(r)$ distributions of ${}^{208}$Pb for the five neutron-skin configurations considered in Table~\ref{tab:rnp}. The upper panel shows $\rho(r)$ corresponding to different $\Delta r_{np}$. The lower panel presents the ratios of $\rho(r)$ for the four nonzero neutron-skin cases to the reference configuration with $\Delta r_{np}=0$, i.e., $\rho(r)\{\Delta r_{np}\neq 0\}/\rho(r)\{\Delta r_{np}=0\}$. Increasing the neutron diffuseness reduces the density in the interior region and redistributes nucleons toward the surface and tail regions while preserving the total nuclear mass. In the lower panel, the ratio ranges approximately from $0.9$ to $1.05$ in the displayed radial range. For compact configurations with smaller or negative $\Delta r_{np}$, the interior density is enhanced compared with the reference case, whereas for more diffuse configurations with positive $\Delta r_{np}$ the interior density is depleted. These density variations provide the geometric baseline for analyzing how neutron-skin-induced modifications of the nuclear surface affect the initial eccentricities and final-state anisotropic flow.
	
	These density distributions are used to sample the initial proton and neutron positions in the improved AMPT-SM model. Once the nucleon coordinates are generated according to Eq.~(\ref{eq:ws}), the model simulates the full dynamical evolution of Pb+Pb collisions. Variations of the initial neutron density can therefore be followed through partonic scattering, quark coalescence, and hadronic interactions, allowing us to examine whether and to what extent their geometric imprints survive in final-state observables.

	\begin{figure}
		\begin{minipage}[c]{0.5\textwidth}
			\subfigure{\includegraphics[width=0.8\textwidth]{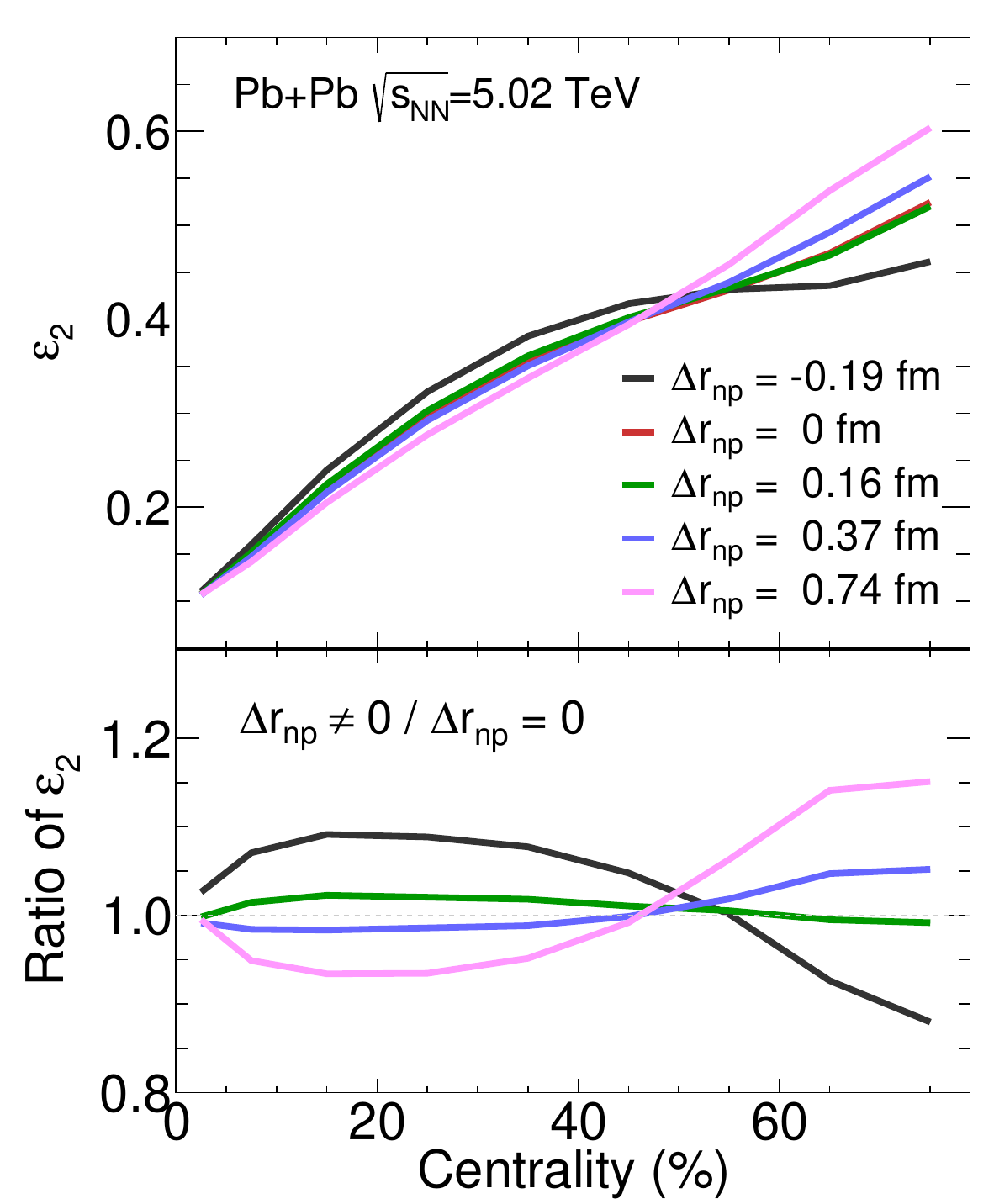}}
		\end{minipage}
		\caption{Upper panel: The centrality dependence of $\varepsilon_2$ in $\rm Pb+Pb$ collisions with different $\Delta r_{np}$. Lower panel: The centrality dependence of the ratio of $\varepsilon_2$ with $\Delta r_{np}$ relative to the reference configuration with $\Delta r_{np}=0$.}
		\label{fig:e2}
	\end{figure}
	
	\begin{figure}
		\begin{minipage}[c]{0.5\textwidth}
			\subfigure{\includegraphics[width=0.8\textwidth]{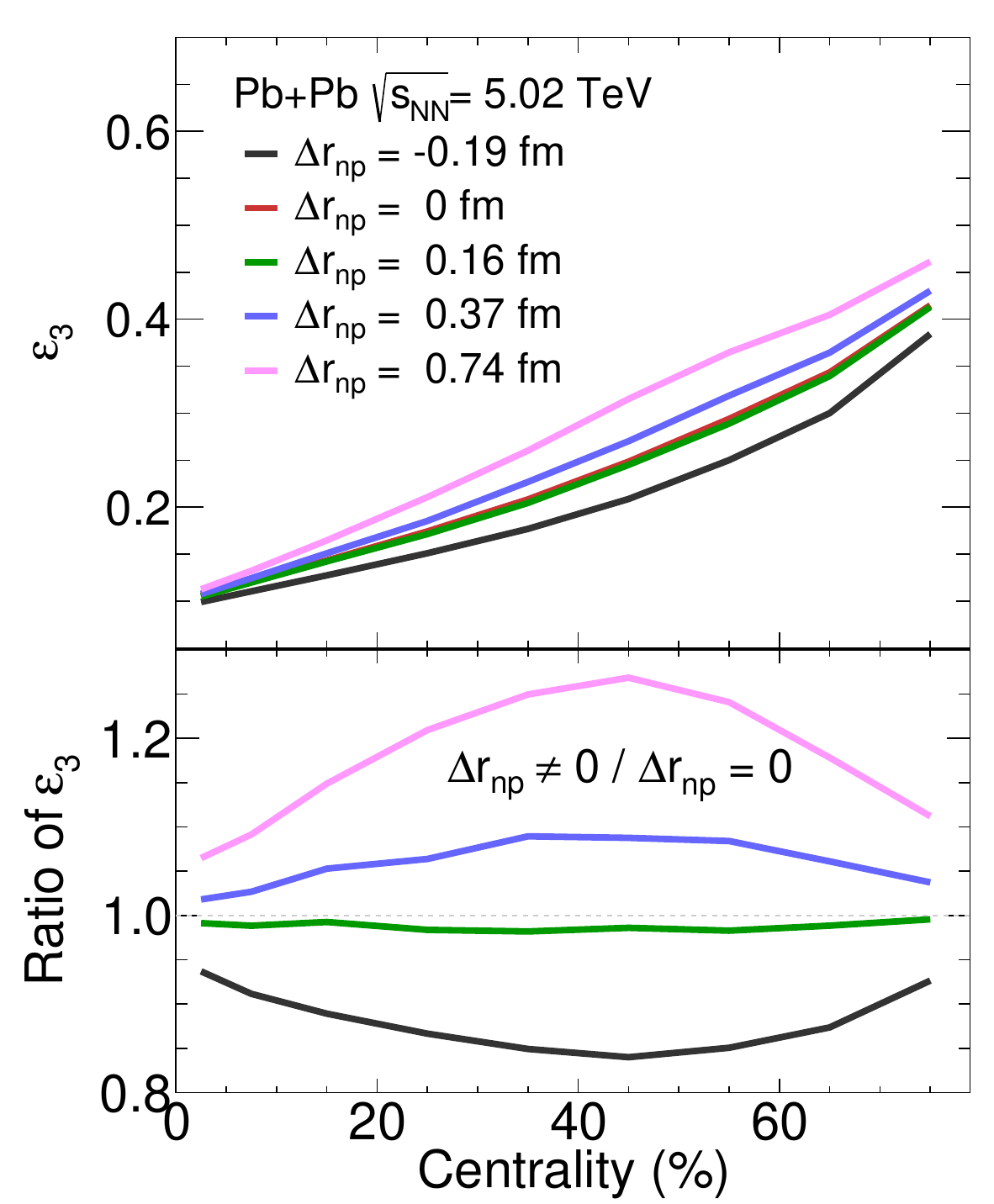}}
		\end{minipage}
		\caption{Upper panel: The centrality dependence of $\varepsilon_3$ in $\rm Pb+Pb$ collisions with different $\Delta r_{np}$. Lower panel: The centrality dependence of the ratio of $\varepsilon_3$ with $\Delta r_{np}$ relative to the reference configuration with $\Delta r_{np}=0$.}
		\label{fig:e3}
	\end{figure}
	
	\begin{figure}
		\begin{minipage}[c]{0.5\textwidth}
			\subfigure{\includegraphics[width=0.8\textwidth]{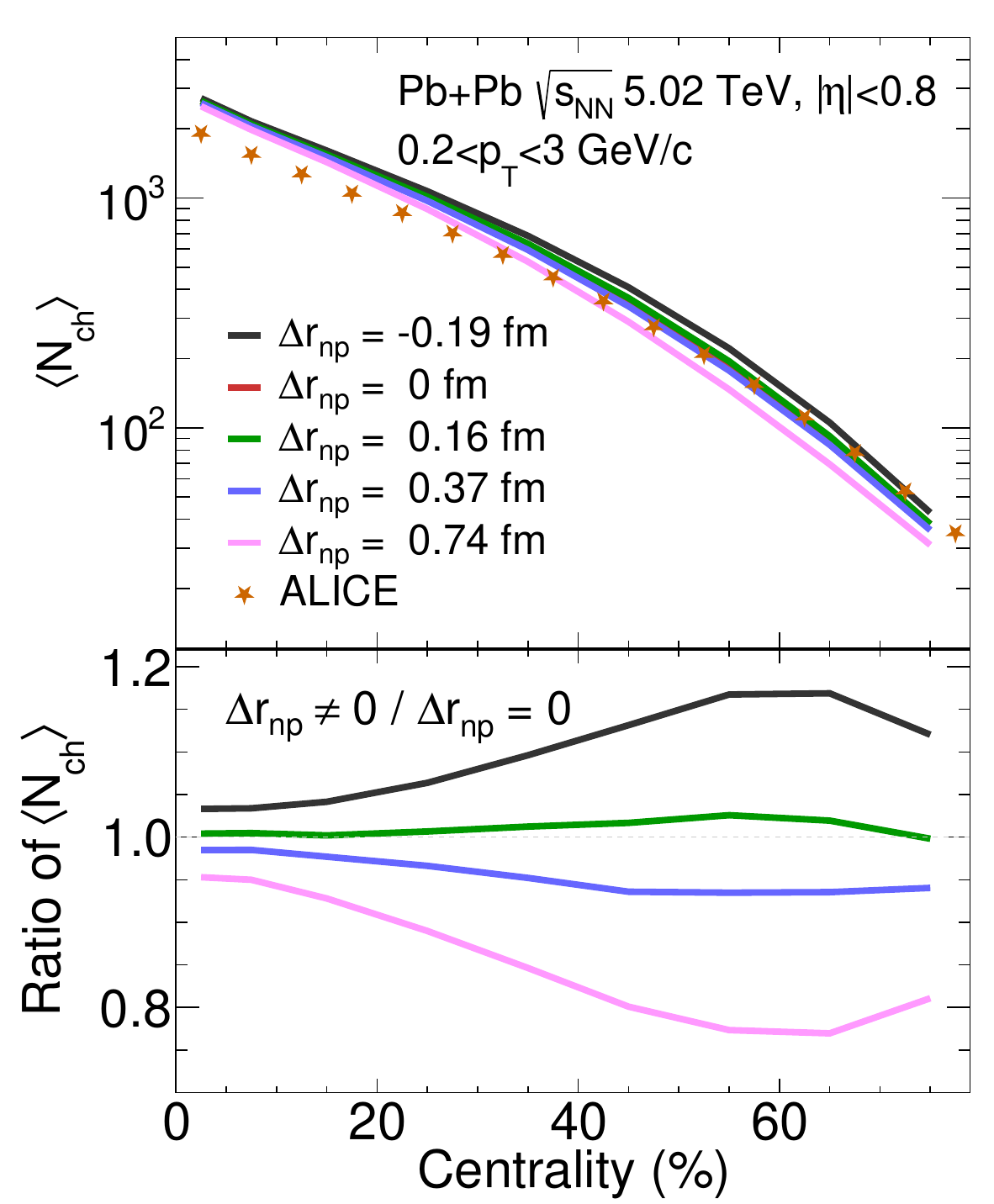}}
		\end{minipage}
		\caption{Upper panel: The centrality dependence of $\left<N_{\rm ch} \right>$ in $\rm Pb+Pb$ collisions with different $\Delta r_{np}$. Lower panel: The centrality dependence of the ratio of $\left<N_{\rm ch} \right>$ with $\Delta r_{np}$ relative to the reference configuration with $\Delta r_{np}=0$.}
		\label{fig:nchcen}
	\end{figure}
	
	\begin{figure}
		\begin{minipage}[c]{0.5\textwidth}
			\subfigure{\includegraphics[width=0.8\textwidth]{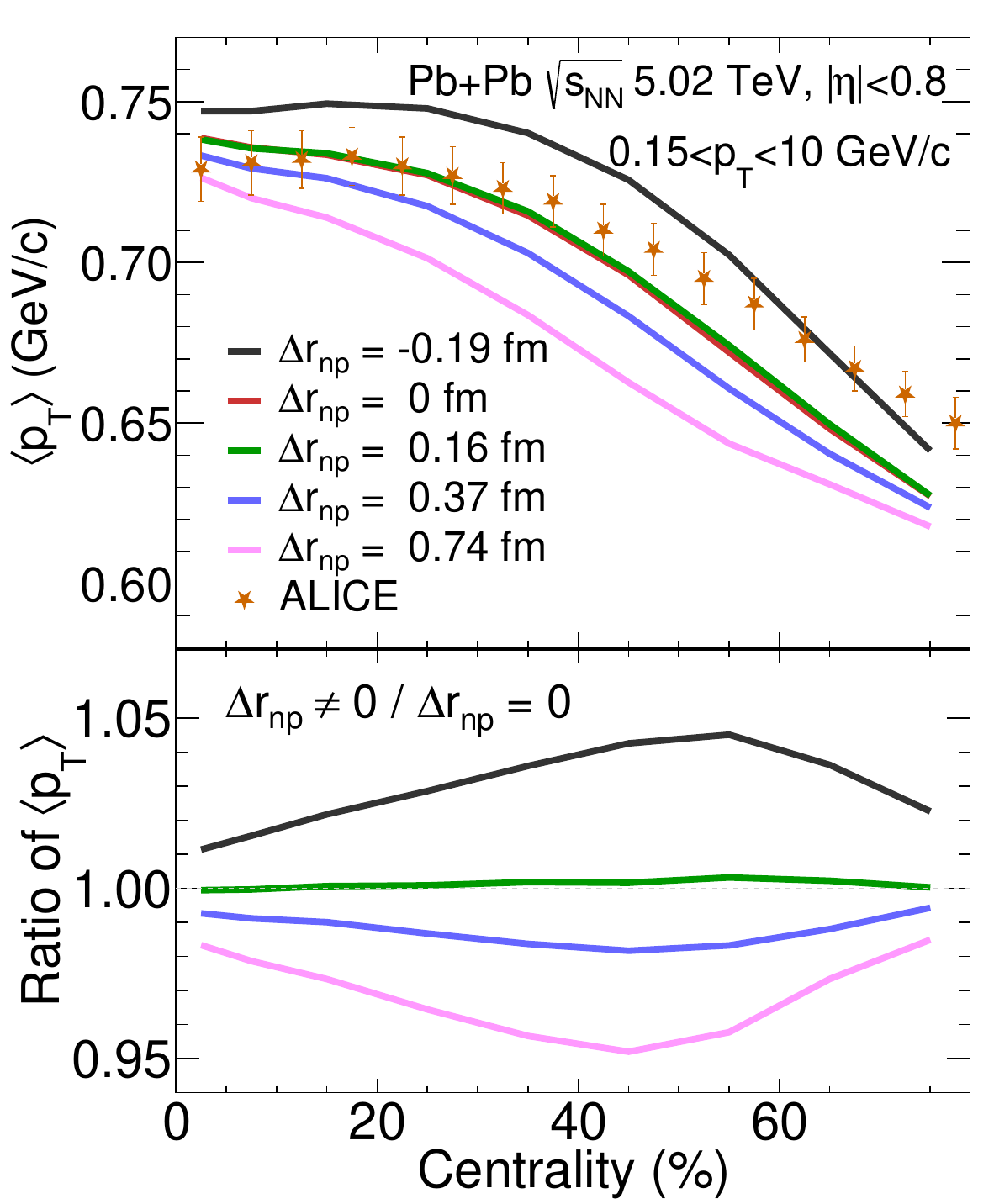}}
		\end{minipage}
		\caption{Upper panel: The centrality dependence of $\left<p_{\rm T} \right>$ in $\rm Pb+Pb$ collisions with different $\Delta r_{np}$. Lower panel: The centrality dependence of the ratio of $\left<p_{\rm T} \right>$ with $\Delta r_{np}$ relative to the reference configuration with $\Delta r_{np}=0$.}
		\label{fig:mpt}
	\end{figure}
	
	\begin{figure}
		\begin{minipage}[c]{0.5\textwidth}
			\subfigure{\includegraphics[width=0.8\textwidth]{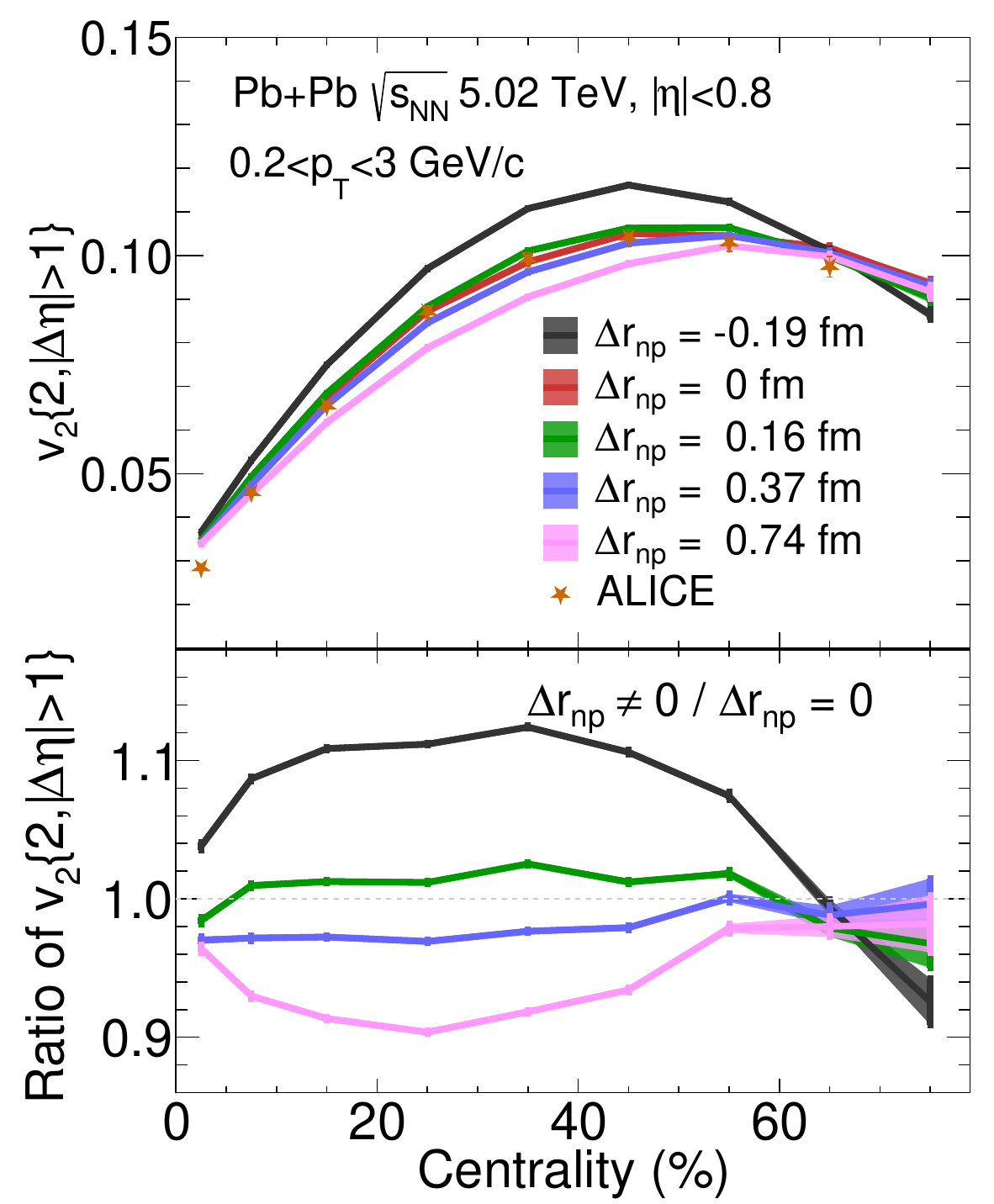}}
		\end{minipage}
		\caption{Upper panel: The centrality dependence of $v_2\{2,|\Delta\eta|>1\}$ in $\rm Pb+Pb$ collisions with different $\Delta r_{np}$. Lower panel: The centrality dependence of the ratio of $v_2\{2,|\Delta\eta|>1\}$ with $\Delta r_{np}$ relative to the reference configuration with $\Delta r_{np}=0$.}
		\label{fig:v2}
	\end{figure}
	
	\begin{figure}
		\begin{minipage}[c]{0.5\textwidth}
			\subfigure{\includegraphics[width=0.8\textwidth]{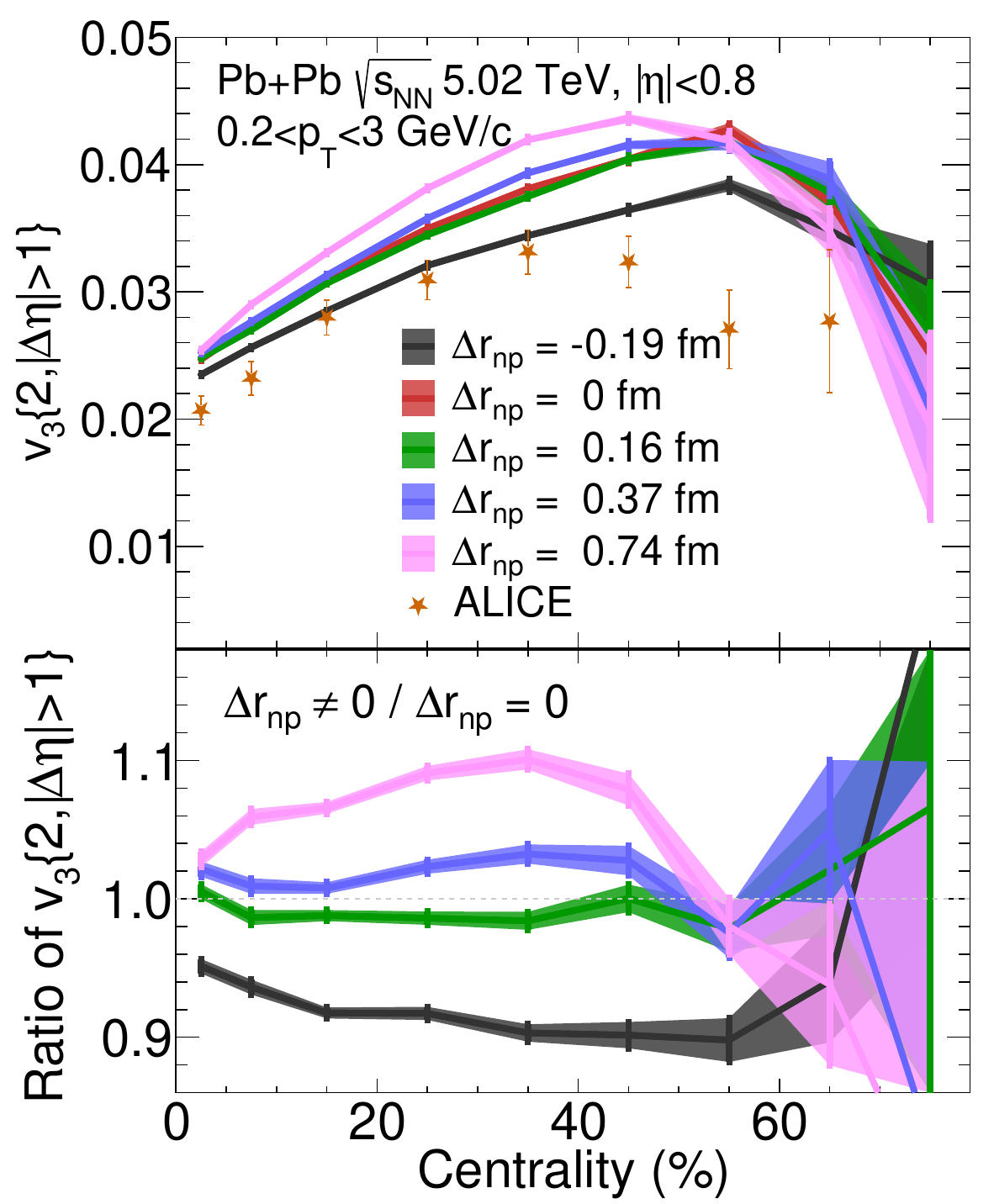}}
		\end{minipage}
		\caption{Upper panel: The centrality dependence of $v_3\{2,|\Delta\eta|>1\}$ in $\rm Pb+Pb$ collisions with different $\Delta r_{np}$. Lower panel: The centrality dependence of the ratio of $v_3\{2,|\Delta\eta|>1\}$ with $\Delta r_{np}$ relative to the reference configuration with $\Delta r_{np}=0$.}
		\label{fig:v3}
	\end{figure}
	
	\begin{figure}
		\begin{minipage}[c]{0.5\textwidth}
			\subfigure{\includegraphics[width=0.8\textwidth]{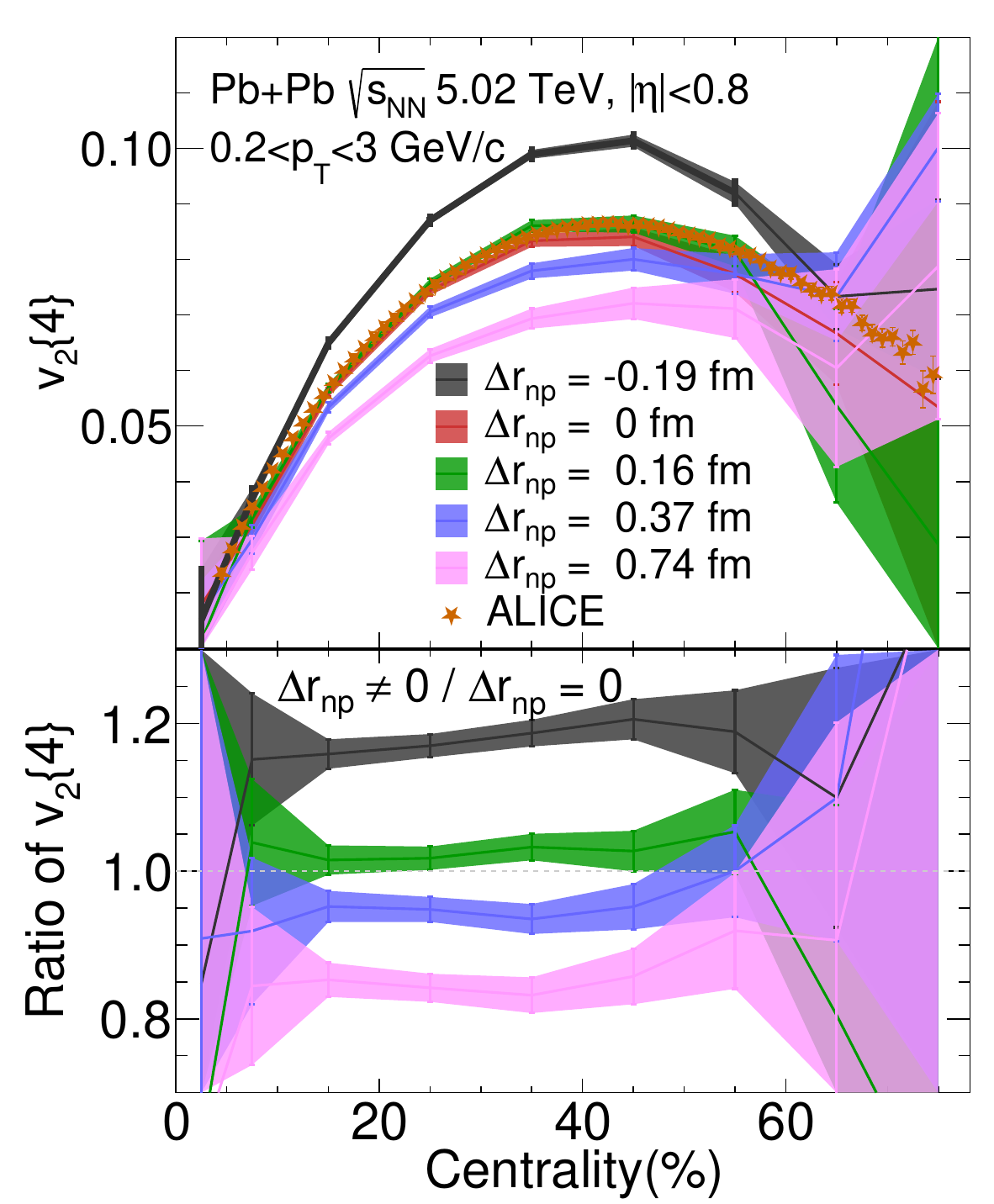}}
		\end{minipage}
		\caption{Upper panel: The centrality dependence of $v_{2}\{4\}$ in $\rm Pb+Pb$ collisions with different $\Delta r_{np}$. Lower panel: The centrality dependence of the ratio of $v_{2}\{4\}$ with $\Delta r_{np}$ relative to the reference configuration with $\Delta r_{np}=0$.}
		\label{fig:v24}
	\end{figure}
	
	\section{\label{sec:results}Results and discussion}
	
	\subsection{Initial geometry}
	
	The centrality dependences of the initial eccentricities $\varepsilon_2$ and $\varepsilon_3$~\cite{Zhao:2024feh} for the five neutron-skin configurations of $^{208}$Pb are shown in Figs.~\ref{fig:e2} and \ref{fig:e3}. $\varepsilon_2$ and $\varepsilon_3$ are displayed in the upper panels, while the lower panels present the ratios of $\varepsilon_n$ for the four nonzero neutron-skin cases to the reference configuration with $\Delta r_{np}=0$, i.e., $\varepsilon_n\{\Delta r_{np}\neq 0\}/\varepsilon_n\{\Delta r_{np}=0\}$ and this convention is adopted for all subsequent ratio plots. In this work, centrality is determined from the final charged-particle multiplicity distribution within $|\eta|<0.8$ and $0.2<p_{\rm T}<2~\mathrm{GeV}/c$, which provides a practical AMPT centrality estimator for comparison with ALICE data~\cite{ALICE:2018ekf}.
	
	The neutron-skin effect on $\varepsilon_2$ is shown in Fig.~\ref{fig:e2}, where a clear ordering among the neutron-skin configurations is observed in central and mid-central collisions. Excluding the reference configuration, more compact nuclei, with smaller or negative $\Delta r_{np}$, result in systematically larger elliptic eccentricities, whereas more diffuse neutron distributions suppress $\varepsilon_2$. This behavior directly reflects the trends of the density profiles shown in Fig.~\ref{fig:rho}, where compact configurations enhance the geometric anisotropy of the overlap region for a given impact parameter. Quantitatively, $\varepsilon_2$ decreases monotonically from the most compact case $\Delta r_{np}=-0.19~\mathrm{fm}$ to the most diffuse one $\Delta r_{np}=0.74~\mathrm{fm}$ over a broad centrality range. In peripheral collisions, however, the ordering is reversed, indicating a transition to a fluctuation-dominated regime in which surface nucleons play a more important role and diffuse density profiles enhance shape irregularities. We further note that the configuration with $\Delta r_{np}=0.16~\mathrm{fm}$ remains very close to the reference case over most of the centrality range, which may be attributed to the fact that both configurations have nearly identical root-mean-square nuclear radii, $\sqrt{\langle r^2 \rangle}\approx 5.53~\mathrm{fm}$.
	
	A qualitatively different but equally systematic trend is observed for the triangular eccentricity $\varepsilon_3$. As shown in Fig.~\ref{fig:e3}, $\varepsilon_3$ increases monotonically with increasing neutron skin over the entire centrality range excluding the reference configuration. This reflects the growing importance of peripheral nucleons in more diffuse nuclei, which amplifies event-by-event spatial irregularities and thus enhances $\varepsilon_3$. The near overlap between the $\Delta r_{np}=0.16~\mathrm{fm}$ and $\Delta r_{np}=0$ cases persists also for $\varepsilon_3$, further supporting the idea that moderate neutron-skin effects may be partially obscured by similarities in the effective initial nuclear geometry resulting from nearly identical nuclear radii.
	
	\subsection{\label{sec:sub1}Final-state observables}
	
	Figure~\ref{fig:nchcen} presents the charged-particle multiplicity $\left<N_{\rm ch} \right>$ as a function of centrality for the five neutron-skin configurations, together with the ALICE data~\cite{ALICE:2018ekf}. Excluding the reference configuration, $\left<N_{\rm ch} \right>$ exhibits a clear and monotonic decrease with increasing neutron skin over the full centrality range. This trend is consistent with the dilution of the nuclear overlap region for more diffuse density profiles, which reduces the average participant density in the transverse plane and consequently lowers the entropy deposition and final-state multiplicity. The relative variation is sizable, with ratios spanning approximately $0.75$ to $1.2$ depending on centrality and neutron-skin thickness.
	Notably, the configuration with $\Delta r_{np}=0.16~\mathrm{fm}$ yields ratios systematically above unity, indicating a modest enhancement compared to the reference geometry. While the model slightly overestimates the multiplicity in the most central events, several configurations reproduce the ALICE data within uncertainties in mid-central and peripheral collisions. 
	Although absolute particle yields may not be perfectly reproduced by the model, ratios among calculations with different density configurations can partially cancel common model dependencies. These ratios indicate that the centrality dependence of $\langle N_{\rm ch}\rangle$ is sensitive to changes in the implemented matter geometry.
	
	The mean transverse momentum $\langle p_T \rangle$, shown in Fig.~\ref{fig:mpt}, follows a similar but weaker trend. Excluding the reference configuration, $\langle p_T \rangle$ decreases monotonically with increasing $\Delta r_{np}$ over the full centrality range, reflecting the reduced transverse pressure gradients associated with broader initial energy-density profiles in nuclei with larger neutron skins. In contrast to the strong effect on $\left<N_{\rm ch} \right>$, the relative variations are small. The $\Delta r_{np}=0.16~\mathrm{fm}$ configuration is almost indistinguishable from the reference case, with ratios confined to $0.95$--$1.05$. This indicates that $\langle p_T \rangle$ is only weakly sensitive to neutron-skin effects. The model reproduces the ALICE data~\cite{ALICE:2018ekf} well in central collisions, while underestimating the measurements in more peripheral events, pointing to the possible relevance of additional dynamical effects in this region.
	
	The centrality dependence of the two-particle flow coefficients $v_2\{2\}$ and $v_3\{2\}$ for the five neutron-skin configurations is presented in Figs.~\ref{fig:v2} and \ref{fig:v3}, together with the ALICE measurements~\cite{ALICE:2018rtz}.
	For $v_2\{2\}$ in Fig.~\ref{fig:v2}, the results decrease monotonically with increasing $\Delta r_{np}$ excluding the reference configuration, with ratios in the range $0.9$--$1.1$, comparable to the relative variation of the initial ellipticity $\varepsilon_2$. This close correspondence indicates that the neutron-skin dependence of $v_2\{2\}$ is largely governed by the transformation of the initial geometry. Among the configurations considered, $\Delta r_{np}=0$ and $0.16~\mathrm{fm}$ provide the best overall description of the data across centrality.
	In Fig.~\ref{fig:v3}, a qualitatively opposite trend is found for $v_3\{2\}$, which increases monotonically with neutron skin, in line with the behavior of $\varepsilon_3$. Although the corresponding ratios remain within $0.9$--$1.1$, the overall variation is smaller than that observed for the initial triangular eccentricity $\varepsilon_3$, indicating that the sensitivity to the neutron skin is diluted during the subsequent transport evolution. In comparison with the data, the most compact configuration, $\Delta r_{np}=-0.19~\mathrm{fm}$, provides the closest description of $v_3\{2\}$.
	
	Figure~\ref{fig:v24} presents the four-particle cumulant elliptic flow $v_2\{4\}$ as a function of centrality. Owing to its strong suppression of nonflow correlations and its direct sensitivity to genuine multiparticle collectivity and event-by-event flow fluctuations, $v_2\{4\}$ offers a stringent test of whether the geometry modifications induced by the neutron skin persist through the full dynamical evolution. A pronounced and systematic dependence on the neutron skin is observed in Fig.~\ref{fig:v24}. Within the nonzero-skin sequence, $v_2\{4\}$ decreases monotonically with increasing $\Delta r_{np}$ over the entire centrality range, with ratios spanning approximately $0.8$--$1.2$. The magnitude of this variation is comparable to that seen for $\varepsilon_2$ and $v_2\{2\}$, indicating that the neutron skin affects not only the mean elliptic response but also the event-by-event structure of the initial geometry that controls flow fluctuations. The comparison with ALICE data~\cite{ALICE:2018rtz} further highlights the discriminating power of $v_2\{4\}$ among the tested density profiles, and the configuration with $\Delta r_{np}=0.16~\mathrm{fm}$ provides the best overall agreement, followed by $\Delta r_{np}=0$, whereas larger neutron skins lead to an increasingly suppressed four-particle cumulant flow. 
	These results indicate that $v_2\{4\}$ is a useful observable for testing neutron-density-induced geometry modifications, although its sensitivity is not unique to $\Delta r_{np}$ and remains correlated with the overall matter geometry.
	
	Using the same kinematic acceptance as in the inclusive-flow analysis, we have also examined charge-dependent pion observables with the same AMPT event samples.
	The yield ratio $N_{\pi^-}/N_{\pi^+}$ and the corresponding flow ratios $v_n^{\pi^-}/v_n^{\pi^+}$ remain consistent with unity and exhibit no discernible dependence on the tested neutron-density configurations within the statistical precision of the calculations. We therefore
	do not include these ratios as separate figures. Their weak sensitivity is expected at LHC midrapidity, where pion production is dominated by pair-produced particles and all the configurations have the same global neutron-to-proton ratio. This supplementary check supports the interpretation that the response of the inclusive observables discussed above is predominantly geometric rather than a direct manifestation of isospin transport. A reliable quantitative study of neutron/proton or mirror-nucleus observables would require a dedicated treatment of isospin-dependent dynamics and, for mirror nuclei, of light-nucleus formation. These extensions are beyond the scope of the present AMPT-SM calculation.
	
	\subsection{Model-to-data comparison and geometric degeneracy}
	\label{subsec:chi2}
	
	\begin{table}[!ht]
		\caption{$\chi^2$ values characterizing the model-to-data agreement for the different neutron-skin configurations in Pb+Pb collisions.}
		\renewcommand\arraystretch{1.5}
		\label{tab:chi}
		\begin{tabular}{c|ccccc}
			\hline\hline
			$\Delta r_{np}$ (fm) & $-0.19$ & $0$ & $0.16$ & $0.37$ & $0.74$ \\
			\hline
			$\chi^{2}$           & ~~100.03~~    & ~~3.43~~ & ~~3.67~~   & ~~11.46~~   & ~~70.89~~   \\
			\hline\hline
		\end{tabular}
	\end{table}
	
	To quantify how current data discriminate among the tested neutron-density profiles in Pb+Pb collisions, we perform a $\chi^2$ analysis to assess the overall agreement between model calculations and experimental measurements. In our baseline analysis, we construct the $\chi^2$ using observables that are most sensitive to neutron-skin variations, namely $\left<N_{\rm ch} \right>$, $v_2\{2\}$, $v_3\{2\}$, and $v_2\{4\}$, and restrict the comparison to the mid-central region $20$--$60\%$, where collective flow dominates and exhibits a clear response to geometric variations. The resulting $\chi^2$ values for the five neutron-skin configurations are summarized in Table~\ref{tab:chi}. For $\Delta r_{np} = -0.19,\ 0,\ 0.16,\ 0.37,$ and $0.74~\mathrm{fm}$, we obtain $\chi^2 = 100.03,\ 3.43,\ 3.67,\ 11.46,$ and $70.89$, respectively.
	Among the five cases, the configuration with $\Delta r_{np}=0$ provides the best overall description of the ALICE data, followed closely by $\Delta r_{np}=0.16~\mathrm{fm}$, while configurations corresponding to a negative neutron skin or a very large positive neutron skin yield substantially larger $\chi^2$ values and are therefore strongly disfavored by the data. The large $\chi^2$ values obtained for these extreme configurations indicate that such neutron-skin implementations lead to initial geometries and subsequent collective dynamics that are less compatible with the measured observables. In this sense, the comparison with data strongly disfavors these extreme scenarios within the present transport-model framework, even though no single observable alone provides a decisive constraint.
	We have also examined the $\chi^2$ values obtained under alternative centrality selections or for individual observables. Although the absolute values of $\chi^2$ change under these alternative choices, the relative ordering among the different neutron-skin configurations remains unchanged, leading to the same qualitative conclusion.
	
	In contrast, the two preferred configurations with $\Delta r_{np}=0$ and $0.16~\mathrm{fm}$ yield very similar $\chi^2$ values, indicating that they correspond to nearly equivalent effective collision geometries in Pb+Pb collisions. This near degeneracy can be attributed, at least in part, to the fact that these two cases are characterized by almost identical nuclear root-mean-square radii, which result in very similar transverse density profiles and overlap geometries. For ${}^{208}$Pb, such moderate variations of the neutron-skin thickness therefore induce only limited changes in the global nuclear size and participant geometry, leading to nearly indistinguishable anisotropic flow patterns within the current experimental precision.
	Our findings indicate that, in large collision systems, the anisotropic flow measurements studied here are primarily influenced by the impact of the neutron skin on the collision geometry and size, and they have limited sensitivity to the fine details of the nuclear surface profile. Thus, the present comparison discriminates against extreme density profiles but does not uniquely determine $\Delta r_{np}$ as an isolated nuclear-structure parameter. Future experimental data with improved precision, or observables with enhanced sensitivity to the nuclear surface, will be needed to further reduce this geometric degeneracy.

	\section{Summary}\label{sec:sum}
	
	We have carried out a systematic study of the sensitivity of anisotropic flow in Pb+Pb collisions at $\sqrt{s_{\mathrm{NN}}}=5.02$~TeV to the neutron skin of $^{208}$Pb within the improved AMPT-SM model. Using the Woods-Saxon density profiles listed in Table~\ref{tab:rnp}, we varied the neutron surface diffuseness to generate several neutron-skin scenarios and traced their impact from the initial nuclear density profiles to the final-state charged-particle production and anisotropic flow. The purpose of this work is not to provide a model-independent extraction of $\Delta r_{np}$, but to test, within a microscopic multiphase transport framework, whether neutron-density-induced geometric modifications can survive partonic scattering, quark coalescence, and hadronic interactions. Our results show that the neutron-skin-induced modifications are systematically reflected in the initial eccentricities and are partially preserved in final-state anisotropic flow. The ordering of the initial eccentricities with the neutron-skin scenario is maintained in the final-state flow observables, including $v_2\{2\}$, $v_3\{2\}$, and especially the four-particle cumulant $v_2\{4\}$. The response of $v_2\{4\}$ indicates that modifications of the participant geometry associated with the neutron density profile survive the subsequent transport evolution and remain encoded in genuine multiparticle correlations. At the same time, the sensitivity of these inclusive flow observables is primarily geometric: they respond to the effective matter distribution, nuclear size, and participant geometry rather than uniquely to $\Delta r_{np}$ alone.
	
	A model-to-data comparison with ALICE measurements shows clear discrimination among the tested density profiles. The most compact configuration and the most diffuse configuration are strongly disfavored within the present AMPT setup, whereas the reference configuration with $\Delta r_{np}=0$ and the moderate-skin configuration with $\Delta r_{np}=0.16~\mathrm{fm}$ provide comparably good descriptions of the data. These two configurations cannot be clearly distinguished with the present charge-inclusive observables, because they generate nearly identical effective collision geometries and root-mean-square nuclear radii. This demonstrates a geometric degeneracy: different neutron-density implementations that produce similar participant geometries can lead to nearly indistinguishable anisotropic flow. Supplementary checks of charge-dependent pion observables show no discernible sensitivity to the tested configurations, further supporting the predominantly geometric origin of the observed response. The discrimination achieved here should therefore be interpreted as model-dependent discrimination among the implemented effective matter geometries, rather than as a unique posterior determination of the neutron-skin thickness of $^{208}$Pb.
	
	This work establishes a microscopic transport-based assessment of how neutron-skin-induced geometric modifications are transmitted to anisotropic flow in Pb+Pb collisions, and clarifies both the sensitivity and the intrinsic limitation of using inclusive flow observables to probe neutron-skin effects. Future progress will benefit from improved experimental precision of multiparticle correlations, a more systematic exploration of nuclear-density parametrizations, observables with enhanced sensitivity to the nuclear surface, and complementary studies in smaller or asymmetric collision systems, which may help to further reduce the remaining geometric degeneracy.

	\begin{acknowledgments}
		We sincerely thank Zi-Wei Lin, Chao Zhang, and Liang Zheng for providing the improved version of the AMPT-SM model and for their substantial contributions to its development and improvement, which were essential for the successful completion of this work. We thank Dr. Chun-Jian Zhang for helpful discussions and Dr. Chen Zhong for maintaining the high-quality performance of Fudan supercomputing platform for nuclear physics. This work is supported by the National Natural Science Foundation of China under Grant No. 12105054 (X.Z.), 
		the National Natural Science Foundation of China under Grants No. 12325507, No. 12547102, and No. 12147101, and the National Key Research and Development Program of China under Grant No. 2022YFA1604900 (G.M.)
		
	\end{acknowledgments}

	\bibliography{refs}

\end{document}